\documentclass[aps,twocolumn,pra,showpacs,floatfix]{revtex4-2}
\usepackage{epsfig}
\usepackage{graphicx}
\usepackage{dcolumn}
\usepackage{amssymb,amsmath}
\usepackage{mathrsfs}
\begin{document}

\title{Electronic bridge  for the $J=0$ state and effects of electrons on frequency  and lifetime of the nuclear clock transition in $^{229}$Th$^{2+}$  ion}

\author{V. A. Dzuba and V. V. Flambaum}

\affiliation{School of Physics, University of New South Wales, Sydney 2052, Australia}

\begin{abstract}
We investigate the electronic bridge (EB) process for excitation of the $^{229}$Th nuclear isomer in Th III ions initially prepared in the metastable electronic state with $J=0$. This state is of particular interest for developing a nuclear clock with suppressed electronic hyperfine interactions. Using a combination of configuration interaction and coupled-cluster methods, we calculate EB transition rates for several excitation schemes involving one or two laser fields. We express the rates in terms of the enhancement factor $R$, defined as the ratio of the EB transition rate to the direct nuclear M1 transition rate. Schemes that return the ion to its initial $J=0$ state have relatively small enhancement factors, $R \sim 1$. However, allowing the final electronic state to differ from the initial state produces a strong resonant enhancement: $R$ reaches $5\times10^3$.
  In addition, we calculate the shift in the nuclear transition frequency caused by changing the electronic state from the ground state to  two excited metastable states. We also examine how the electronic state affects the nuclear transition lifetime and calculate the probabilities of electron transitions induced by nuclear decay. The EB mechanism shortens the nuclear decay lifetime by a factor of 2.6 when the electrons are initially in the ground state and by a factor of 6.85 when they are in the metastable $J=0$ state.
\end{abstract}

\maketitle

\section{Introduction}

The nucleus of $^{229}\mathrm{Th}$ has a unique feature: a low-lying isomeric state~\cite{Th1} that can decay to the nuclear ground state via an M1 transition. This transition has been proposed as the basis for a nuclear clock~\cite{Th2}, with a potential relative precision $\sim 10^{-19} - 10^{-20}$~\cite{Kuzmich,ThV}. The thorium nuclear clock is also expected to provide high sensitivity to possible variations of fundamental physical constants, violation of fundamental symmetries  and  low mass scalar and pseudo-scalar (axion) fields ~\cite{Flambaum,Flambaum2016,Stadnik,Perez}.

The energy of the nuclear transition has been measured to be $\omega_N = 8.355733(2)_{\rm stat}(10)_{\rm sys}$~eV (67393 cm$^{-1}$)~\cite{Th-wn} and $2,020,407,384,335(2)$ kHz~\cite{Th-wn1} for $^{229}\mathrm{Th}$ embedded in solids. Higher accuracy is expected to be achievable with nuclear clocks based on trapped ions~\cite{Kuzmich,ThV}.

Direct nuclear excitation by a laser field is challenging because the nuclear M1 transition amplitude is suppressed by approximately five orders of magnitude compared with typical atomic E1 transition amplitudes. This suppression arises from the small nuclear magneton, $\mu_N=e\hbar/(2m_p)$, which characterizes nuclear magnetic moments, compared with the characteristic atomic electric-dipole scale $ea_0$. In particular, the ratio $\mu_N/(ea_0c)$ is of the order of $10^{-6}$, where $c$ is the speed of light. The electronic bridge (EB) process offers a possible way to enhance the nuclear excitation rate. In this process, atomic electrons are excited by the laser field, and the subsequent electronic transition transfers energy to the nucleus through the hyperfine interaction~\cite{Tkalya1,Tkalya2}.

Various EB mechanisms for excitation and decay of the nuclear isomer have been investigated in calculations for Th II, Th III, and Th IV; see, for example, Refs.~\cite{EB1,EB2,EB3,EB4,EB5,EB6,EB7,Thii,Thiii}. In particular, Ref.~\cite{Thii} proposed using the EB process for resonant nuclear excitation in Th II.

The details of the EB process depend strongly on the electronic configuration of the ion. In our previous studies of Th II~\cite{Thii} and Th III~\cite{Thiii}, the ions were assumed to be initially in their electronic ground states.Recently, Maurya \emph{et al.}~\cite{Maurya} proposed using a metastable state of Th III with total electronic angular momentum $J=0$. This choice eliminates  hyperfine shifts associated with nonzero electronic angular momentum.

In this work, we investigate the EB process for the proposed configuration using the same approach as in our previous study of Th~II and Th~III~\cite{Thii,Thiii}. We start with relativistic Hartree–Fock (RHF) calculations for the closed-shell Th~V core to generate the self-consistent potential and the single-electron basis states. We then use a combination of the configuration-interaction and single-double coupled-cluster methods (CI+SD)~\cite{CI+SD} to calculate the two-electron valence states of Th~III. Transition amplitudes are calculated using the time-dependent Hartree–Fock method~\cite{TDHF}, which is equivalent to the random-phase approximation (RPA).

\section {States of interest for the electron bridge  processes}


Detailed large-scale calculations of the energy levels and other electronic properties of Th and its ions were presented in Ref.~\cite{Safronova2014}. Our calculated energies are in good agreement with the results of that work. We do not provide a detailed comparison, since the focus of the present study is the EB process, which was not considered in Ref.~\cite{Safronova2014}.

Lifetime of the metastable electronic state  with $J=0$ was calculated in Ref.~\cite{Safronova2014} to be 95.55~s. Its decay is dominated by the E2 transition to the second-lowest state, $6d^2,{}^3F_2$, at 63.27~cm$^{-1}$. Our calculated lifetime of 100~s is in good agreement with the result of Ref.~\cite{Safronova2014}.



\begin{figure}[tb]
	\epsfig{figure=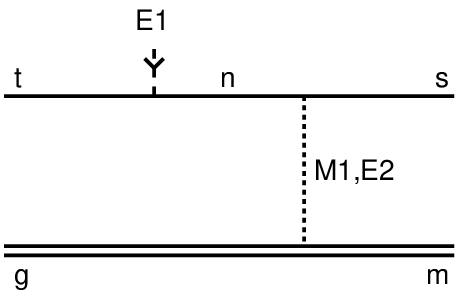,scale=0.6}
	\caption{Diagram for nuclear excitation via the electronic bridge  process in Th~III. It is assumed that the atom is initially in the metastable state 
	S0 (see Table~\ref{t:ener}) which has $J=0$. Then an excited odd state t with $J=1$ is populated by the first laser with frequency 
	$\hbar\omega_1 = E_{\rm t} - E_{\rm S1}$. The second  laser frequency is tuned to excite nucleus into the isomeric state.   
	The E1 operator represents the interaction with the second laser, which has a frequency $\hbar\omega_2 = \hbar\omega_N + E_{\rm s} - \hbar\omega_1$. Here, $n$ and $s$ denote the intermediate and final electronic states, respectively, while $g$ and $m$ correspond to the ground and isomeric nuclear states. The dotted line indicates the hyperfine interaction between atomic electrons and the nucleus.}
	\label{f:eb}
\end{figure}

Calculated and experimental energy levels and Landé $g$-factors for the states of interest are presented in Table~\ref{t:ener}. The state labels correspond to those used in Fig.~\ref{f:eb}.

We assume that the ion is initially in the metastable even-parity state with $J=0$, denoted by S0 in Table~\ref{t:ener}. The ion is first excited by a laser to an odd-parity state $t$ with $J=1$. A second laser then excites the ion to a virtual state $n$, which decays to a final electronic state $s$, while simultaneously exciting the nucleus to the isomeric state $m$ through the hyperfine interaction (HFI). Different choices of the states $t$ and $s$ correspond to different EB schemes, as listed in Table~\ref{t:betan}. The laser frequencies are chosen to satisfy
\[
\hbar\omega_1=E_t-E_{\rm S0},
\qquad
\hbar\omega_2=\hbar\omega_N+E_s-E_t.
\]
Here, $\omega_N$ is the nuclear transition frequency. The diagram in Fig.~\ref{f:eb} represents a sum over the virtual intermediate states $n$, several examples of which are listed in Table~\ref{t:ener}.

The state N6 is of particular importance. When the final electronic state is S1, the EB process is strongly enhanced by the small energy denominator associated with N6,
\[
\Delta=E_{\rm N6}-E_{\rm S1}-\hbar\omega_N
=-72~\mathrm{cm}^{-1}.
\]

A single-laser EB scheme can also be considered. In this case, the state $t$ in Fig.~\ref{f:eb} is identified with the initial state S0, and the final electronic state $s$ has odd parity. Examples of such final states, labelled S2, S3, S4 are listed in Table~\ref{t:ener}.

\begin{table} 
  \caption{\label{t:ener}States of interest for the electronic bridge process in Th~III which
  starts at the metastable state S1 and ends in S1 or S2. Notations T, N, S correspond
  to lines t, n, s on the diagram of Fig.~\ref{f:eb}.}
\begin{ruledtabular}
\begin{tabular}   {lll c rl rl}
\multicolumn{1}{l}{State}&
\multicolumn{1}{l}{Confi-}&
\multicolumn{1}{l}{Term}&
\multicolumn{1}{c}{$J$}&
\multicolumn{2}{c}{NIST~\cite{NIST}}&
\multicolumn{2}{c}{This work}\\
\multicolumn{1}{c}{name}&
\multicolumn{1}{c}{guration}&
&&\multicolumn{1}{c}{Energy}&
\multicolumn{1}{c}{Land\'{e}}&
\multicolumn{1}{c}{Energy}&
\multicolumn{1}{c}{Land\'{e}}\\
&&&&\multicolumn{1}{c}{[cm$^{-1}$]}&&
\multicolumn{1}{c}{[cm$^{-1}$}]& \\
\hline
T1 & $5f6d$ & $^1$P$^{\rm o}$     & 1 &  20710.94 &        &  21630 &  1.0140 \\
T2 & $6d7p$ &                     & 1 &  39280.86 &  0.900 &  38184 &  0.9775 \\
T3 & $7s7p$ & $^3$P$^{\rm o}$     & 1 &  45063.97 &  1.120 &  42880 &  1.0363 \\
T4 & $6d7p$ & $(3/2,3/2)^{\rm o}$ & 1 &  50992.57 &  1.22  &  48884 &  1.2009 \\
   &        &               &     &          &        &        &         \\
N1 & $6d7s$ & $^3$D         &   1 &   5523.88 &  0.50  &   3935 &  0.5000 \\
N2 & $6d^2$ & $^3$P         &   1 &   7875.82 &  1.50  &   5958 &  1.4999 \\
N3 & $5f^2$ & $^3$P         &   1 &  30402.85 &  1.494 &  31525 &  1.4996 \\
N4 & $5f7p$ & (5/2,3/2)     &   1 &  44603.10 &  0.495 &  43915 &  0.5004 \\
N5 & $6d8s$ & $^3$D         &   1 &  81706.37 &        &  81416 &  0.5312 \\
N6 & $6d7d$ & (3/2,3/2)     &   1 &  83358.66 &        &  82952 &  0.7530 \\ 
N7 & $6d7d$ &               &   1 &  85213.26 &        &  85009 &  1.5021 \\
   &        &               &     &           &        &        &         \\
S0 & $6d^2$ & $^3$P         &   0 &   5090.05 &  0.0   &   5282 &  0.0000 \\
   &        &               &     &           &        &        &         \\
S1 & $6d7s$ & $^1$D         &   2 &  16037.64 &  1.00  &  16188 &  1.0377 \\
   &        &               &     &           &        &        &         \\
S2 & $5f6d$ & $^3$P$^{\rm o}$  &   1 &  11123.17 &  1.352  &  11623 &  1.2396 \\
S3 & $5f6d$ & $^3$P$^{\rm o}$  &   0 &  11232.61 &  0         &  11788 &  0 \\
S4 & $5f6d$ & $^3$P$^{\rm o}$  &   2 &  13208.21 &  1.432  &  14128 &  1.3881 \\

\end{tabular}			
\end{ruledtabular}
\end{table}

\section{Calculations for the electronic bridge processes starting from  the metastable state with $J=0$}

\label{s:EB}

In calculating the transition rates for the EB process we follow approach very similar to what we used in our previous works~\cite{Thii,Thiii}.
The rate of an induced excitation  from state $b$ to state $a$ $W_{ba}^{\rm in}$ can be calculated using the rate of a spontaneous transition $a \rightarrow b$ $W_{ab}$  \cite{Sobelman}:
\begin{equation}\label{e:ab-ba}
 W_{ba}^{\rm in} = W_{ab}\frac{4\pi^3c^2}{\omega^3} I_{\omega}.
 \end{equation}
 Here, $I_{\omega}$ is the spectral intensity of the incident isotropic, unpolarized radiation at the transition frequency $\omega$. 
 The index $b$ corresponds to the state with the nucleus in its ground state and the electron in an excited state, while $a$ corresponds to the state with the nucleus excited and the electron in its ground or a low-lying excited state.
The rate of a spontaneous transition via electronic bridge process is given by \cite{EB1}
 \begin{equation}\label{e:geb}
 W_{ab} = \frac{4}{9}\left(\frac{\omega}{c}\right)^3 \frac{|\langle I_g||M_k||I_m \rangle|^2}{(2I_m+1)(2J_t+1)}G_2^{(k)},
 \end{equation}
Here, $\omega$ is the frequency of the laser field driving the nuclear excitation 
(line E1 on Fig.~\ref{f:eb}) 
Factor $G_2$ in (\ref{e:geb}) depends on electrons only. It corresponds to upper line of Fig.~\ref{f:eb}. It is given by
\begin{equation}\label{e:g2a}
G_2^{(k)} = \sum_n \frac{1}{2J_n+1}\left[\frac{\langle s || T_k || n\rangle \langle n||D|| t \rangle}{\omega_{ns} - \omega_N}\right]^2.
\end{equation}
Here $D$  is the electric dipole operator (E1),  $T_k$ is the electron part of the hyperfine interaction operator (magnetic dipole (M1) for $k=1$ and electric quadrupole (E2) for $k=2$).
In the present work we include only magnetic dipole contribution. In most cases electric quadrupole contribution is small and can be neglected.
It might be relatively important in special cases when the E2  contribution is strongly enhanced due to resonance situation but magnetic dipole channel is closed (e.g., $\Delta J =2$). However, E2 does not significantly affect our results for the total widths.

It is convenient to present the results in terms of dimensionless ratios ($\beta$) of EB  transition rates to the nuclear transition rates.
\begin{equation}\label{e:beta}
\beta=\Gamma^{(1)}_{\rm EB}/\Gamma_N(M1).
\end{equation}
Here $\Gamma^{(k)}$ is given by (\ref{e:geb}) ($\Gamma^{(k)} \equiv W_{ab}$). The parameter $\beta$  can be expressed in terms of the electronic factor  $G_2^{(1)}$, (Eq.~(\ref{e:g2a}))~\cite{EB4}
\begin{equation}\label{e:beta1}
\beta= \left(\frac{\omega}{\omega_N}\right)^3\frac{G_2^{(1)}}{3(2J_s+1)}, 
\end{equation}

\begin{table} 
\caption{\label{t:betan} Variants of the electronic bridge  process with resonance nuclear excitation. 
  Notations t and s refer to the diagram on Fig.~\ref{f:eb}. State names Ti, Si correspond
  to the states in Table~\ref{t:ener}.
$\omega_1$ and $\omega_2$ are the frequencies of the first and second excitations respectively,
$\hbar\omega_1=E_{\rm f} - E_{\rm s1}$, $\hbar\omega_2=\hbar\omega_N+E_{\rm s}-\hbar\omega_1$.
  In first four variants the initial and final states are the same, the metastable state
  with $J=0$. Here $\beta$ is small ($\beta \sim 1$) due to absence of resonance contribution.
  In next four variants the final state S2 has been chosen to get large enhancement due
  to the term with small energy denominator, $\Delta_n = E_{\rm N6} -\hbar\omega_N - E_{\rm S2} = -72 \ {\rm cm}^{-1}$.
  Last three variants correspond to single excitation by a laser with $\hbar\omega_2 = \hbar\omega_N - E_{\rm S0} + E_{\rm s}$.
  }

\begin{ruledtabular}
\begin{tabular}{cccccc rr}

& t & s  & $\omega_1$ & $\omega_2$  & $\beta$ \\
&   &    & [cm$^{-1}$] & [cm$^{-1}$] & \\
\hline

1 & T1 & S0 & 15621 & 51773 & 1.69 \\
2 & T2 & S0 & 34191 & 33203 & 0.82 \\
3 & T3 & S0 & 39974 & 27420 & 0.89 \\
4 & T4 & S0 & 45903 & 21490 & 0.39 \\
  &    &    &       &       &      \\
5 & T1 & S1 & 15621 & 62721 & 1.1[+3] \\
6 & T2 & S1 & 34191 & 44150 & 5.0[+3] \\
7 & T3 & S1 & 39974 & 38368 & 2.2[+2] \\
8 & T4 & S1 & 45903 & 32438 & 3.7[+3] \\
  &    &    &       &       &      \\
 9  & S0 & S2  &  & 76479 & 1.2 \\
10 & S0 & S3  &  & 76692 & 1.6 \\
11 & S0 & S4  &  & 76883 & 0.038 \\
\end{tabular}			
\end{ruledtabular}
\end{table}
The results of the calculations are presented in Table~\ref{t:betan}. We consider three distinct groups of EB processes. In the first group, the initial and final electronic states are the same: the metastable state S0 with $J=0$. 
 However, it does not provide an enhancement of the EB rate, with $\beta\sim 1$, due to the absence of resonant contributions.

As demonstrated in our previous works~\cite{Thii,Thiii}, a resonance can be achieved by an appropriate choice of the final electronic state $s$. In this case, the resonance condition is
\[
E_n-E_s\approx\hbar\omega_N.
\]
However, a virtual state $n$ satisfying this condition does not necessarily exist for every choice of the final state $s$.

An advantage of this EB scheme over direct nuclear excitation is that it allows the use of optical lasers. The laser frequencies are determined by the choice of the intermediate state $t$. Four possible choices of this state are considered in Table~\ref{t:betan}.

The second group of EB processes (versions 5–8 in Table~\ref{t:betan}) exhibits strong enhancement due to a resonant contribution with a small energy denominator,
\[
\Delta_n=E_{\rm N6}-\hbar\omega_N-E_{\rm S1} = -72\ \mathrm{cm}^{-1}.
\]
Depending on the choice of the intermediate state $t$, the enhancement can reach nearly four orders of magnitude. The main challenge in this scheme is returning the ion to the initial state S0. 
  Note that S0 state appears as final state in one of the decay channels of the S1 state: S1 $\rightarrow \ ^3$D$^{\rm o}_1, \ E=7921 \ {\rm cm}^{-1} \ \rightarrow$ S0.
The channel is not dominating. The decay of the S1 state is dominated by electric dipole transition to the $^3$F$_2$ state at $E=510.75 \ {\rm cm}^{-1}$.
However, the decay branch to S0 could potentially be enhanced by laser-assisted stimulated emission.

The final three variants in Table~\ref{t:betan} are single-laser EB processes. There is no intermediate state; instead, the initial state S0 plays this role. The final electronic states $s$ have odd parity.

In version 9, the state S0 is one of the decay channels of the state S2. There are six E1 decay channels from S2 to lower-lying even-parity states. The decay is dominated by the E1 transition to the lowest even-parity state, with $J=2$ and $E=63.27~\mathrm{cm}^{-1}$, which accounts for 33\% of the total decay rate. The decay branch to the metastable state S0 accounts for only about 3\% of the total rate. However, this transition can be enhanced by laser stimulation at its resonant frequency. With sufficiently high laser intensity, this stimulated transition could become the dominant decay channel.

\section{Dependence of the nuclear transition lifetime on the states of electrons and spectrum of emitted photons.}
 
\begin{figure}[tb]
	\epsfig{figure=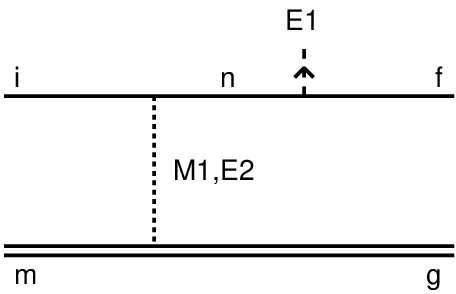,scale=0.6}
	\caption{Electronic bridge contribution to the decay rate of the isomeric nuclear state $m$. Index $i$ stands for the initial atomic state, $n$ stands for intermediate states, $f$ stands for final odd-parity states. }
	\label{f:decay}
\end{figure}

The calculations are similar to those for EB excitation. We follow the approach used in our previous work~\cite{Thii}. The EB contribution to the decay of the nuclear excited state is illustrated by the diagram in Fig.~\ref{f:decay}. The calculation requires summation over all possible intermediate electronic states $n$ and final electronic states $f$.

As before, it is convenient to present the results in terms of dimensionless ratios of EB  transition rates to the nuclear transition rate.
Note, that summation over final states $f$ leads to summation of terms similar to $\beta_{M1}$ (\ref{e:beta1})  with different values of $\omega$: $\hbar\omega \equiv \hbar\omega_f = \hbar\omega_N - E_f +E_i$. Therefore, we introduce new variable $R$ 
\begin{equation}\label{e:R}
R = \sum_f \beta(\omega_f).
\end{equation}

The results of the calculations of the individual terms in Eq.~(6) are presented in Table~III for the metastable initial ionic state with $J=0$ and in Table~IV for the initial ionic ground state. Only contributions satisfying $\beta>0.01R$ are included in the tables.

The values of $R$, obtained by summing the entries in the last columns of the tables, are $R=5.85$ for the metastable initial state and $R=1.60$ for the initial ground state. The latter value is in reasonable agreement with a rough estimate of $R \sim0.7$ reported in Ref.~\cite{Thiii}. That estimate was based on the dominant decay channel and calculated energies of the electronic states of the ion. In contrast, our  present calculation includes all decay channels and uses  more accurate experimental energy levels.

The Tables also give the spectra of emitted photons. Note that frequencies of emitted photons are determined via
experimental values:  $\hbar\omega_f = E_i +\hbar\omega_N - E_f$. 
Since the initial-state energy $E_i$ and the nuclear transition frequency $\omega_N$ are fixed, the emitted photon frequency depends on the final state $f$. The resulting photon frequencies are therefore determined accurately. In contrast, the relative intensities, represented by $\beta_f$, are calculated theoretically. They are sensitive to the configuration mixing and  should be regarded as estimates.
          
\begin{table} 
\caption{\label{t:phot}
The spectrum of emitted photons in the EB-assisted decay of nuclear isomeric state
while ion is in the metastable state $6d^2 \ ^3$P$_0$ at $E=5090$~cm$^{-1}$.
The frequency of the photon is determined by final state via $\hbar\omega_{\rm ph} =
E_i +\hbar\omega_N - E_f$, where $E_i = 5090$~cm$^{-1}$ is the energy of the initial state,
$\hbar\omega_N = 67379$~cm$^{-1}$ is the nuclear frequency. Relative intensities  are given
by the rate coefficients $\beta_f$; $R = \sum_f \beta(\omega_f)=5.85$ gives the total relative contribution of the electronic bridge processes to the nuclear decay probability.
}
\begin{ruledtabular}
\begin{tabular}{ll c rrr}
\multicolumn{4}{c}{Final state}&
\multicolumn{2}{c}{Emitted photon}\\
        &            &$J$& $E_f$       & $\hbar\omega_{\rm ph}$  & $\beta$ \\
        &            &   & [cm$^{-1}$] & [cm$^{-1}$] & \\
\hline
      &              &   0 &    11232.61 &   61250.73  &  0.302 \\
7s.7p &   3P*        &   0 &    42259.71 &   30223.63  &  0.427 \\
      &              &   0 &    51745.37 &   20737.97  &  0.082 \\
5f.6d &   3D*        &   1 &     7921.08 &   64562.26  &  0.893 \\
5f.6d &   3P*        &   1 &    11123.17 &   61360.17  &  0.398 \\
5f.6d &   1P*        &   1 &    20710.94 &   51772.40  &  0.602 \\
      &              &   1 &    39280.86 &   33202.48  &  0.406 \\
7s.7p &   3P*        &   1 &    45063.97 &   27419.37  &  0.356 \\
      &              &   1 &    50992.57 &   21490.77  &  0.190 \\
6d.7p &   (5/2,3/2)* &   1 &    53939.27 &   18544.07  &  0.081 \\
5f.6d &   3F*        &   2 &      510.75 &   71972.59  &  0.446 \\
5f.7s &   3F*        &   2 &     3181.50 &   69301.84  &  0.083 \\
5f.6d &   *          &   2 &     6288.22 &   66195.12  &  0.388 \\
5f.6d &   3D*        &   2 &    10180.76 &   62302.58  &  0.269 \\
5f.6d &   3P*        &   2 &    13208.21 &   59275.13  &  0.334 \\
6d.7p &   (3/2,1/2)* &   2 &    37280.23 &   35203.11  &  0.110 \\
6d.7p &   (5/2,1/2)* &   2 &    44088.28 &   28395.06  &  0.063 \\
6d.7p &   (3/2,3/2)* &   2 &    47679.97 &   24803.37  &  0.195 \\
6d.7p &   *          &   2 &    53151.57 &   19331.77  &  0.079 \\
7s.7p &   3P*        &   2 &    55399.07 &   17084.27  &  0.146 \\  
\end{tabular}			
\end{ruledtabular}
\end{table}
\begin{table} 
\caption{\label{t:photgs}
The spectrum of emitted photons in the EB-assisted decay of nuclear isomeric state
while ion is in the ground state $5f6d \ ^3$H$^{\rm o}_4$.
Relative intensities  are given
by the rate coefficients $\beta_f$; $R = \sum_f \beta(\omega_f)=1.60$ gives the total relative contribution of the electronic bridge processes to the nuclear decay probability.
}
\begin{ruledtabular}
\begin{tabular}{ll c rrr}
\multicolumn{4}{c}{Final state}&
\multicolumn{2}{c}{Emitted photon}\\
        &            &$J$& $E_f$       & $\hbar\omega_{\rm ph}$  & $\beta$ \\
        &            &   & [cm$^{-1}$] & [cm$^{-1}$] & \\
\hline
\multicolumn{6}{c}{$J_n=3$} \\
5f.6d &   3F*        &   2 &      510.75 &   66882.59  &  0.023 \\
5f.7s &   3F*        &   2 &     3181.50 &   64211.84  &  0.014 \\
5f.6d &   *          &   2 &     6288.22 &   61105.12  &  0.007 \\
5f.6d &   3D*        &   2 &    10180.76 &   57212.58  &  0.007 \\
6d.7p &   (3/2,1/2)* &   2 &    37280.23 &   30113.11  &  0.034 \\
6d.7p &   (5/2,1/2)* &   2 &    44088.28 &   23305.06  &  0.009 \\
6d.7p &   (3/2,3/2)* &   2 &    47679.97 &   19713.37  &  0.004 \\
5f.7s &   *          &   3 &     2527.09 &   64866.25  &  0.019 \\
5f.6d &   *          &   3 &     4826.82 &   62566.52  &  0.020 \\
5f.6d &   3G*        &   3 &     5060.54 &   62332.80  &  0.020 \\
5f.7s &   *          &   3 &     7500.60 &   59892.74  &  0.028 \\
      &              &   3 &    10741.15 &   56652.19  &  0.013 \\
      &              &   3 &    44464.95 &   22928.39  &  0.011 \\
      &              &   3 &    49980.67 &   17412.67  &  0.003 \\
5f.6d &   3H*        &   4 &        0.00 &   67393.34  &  0.004 \\
5f.6d &   *          &   4 &     3188.30 &   64205.04  &  0.020 \\
5f.7s &   3F*        &   4 &     6310.80 &   61082.54  &  0.046 \\
5f.6d &   3G*        &   4 &     8141.74 &   59251.60  &  0.033 \\
5f.6d &   *          &   4 &     8980.55 &   58412.79  &  0.017 \\ 
\multicolumn{6}{c}{$J_n=4$} \\
5f.7s &   *          &   3 &     2527.09 &   64866.25  &  0.010 \\
5f.6d &   *          &   3 &     4826.82 &   62566.52  &  0.022 \\
5f.6d &   3G*        &   3 &     5060.54 &   62332.80  &  0.006 \\
5f.7s &   *          &   3 &     7500.60 &   59892.74  &  0.009 \\
      &              &   3 &    10741.15 &   56652.19  &  0.008 \\
5f.6d &   1F*        &   3 &    15453.41 &   51939.93  &  0.006 \\
      &              &   3 &    44464.95 &   22928.39  &  0.002 \\
      &              &   3 &    49980.67 &   17412.67  &  0.002 \\
5f.6d &   3H*        &   4 &        0.00 &   67393.34  &  0.004 \\
5f.6d &   *          &   4 &     3188.30 &   64205.04  &  0.010 \\
5f.7s &   3F*        &   4 &     6310.80 &   61082.54  &  0.011 \\
5f.6d &   3G*        &   4 &     8141.74 &   59251.60  &  0.011 \\
5f.6d &   *          &   4 &     8980.55 &   58412.79  &  0.007 \\
5f.6d &   3H*        &   5 &     4489.64 &   62903.70  &  0.014 \\
5f.6d &   3G*        &   5 &    11276.80 &   56116.54  &  0.025 \\
5f.6d &   1H*        &   5 &    19009.90 &   48383.44  &  0.017 \\  
\multicolumn{6}{c}{$J_n=5$} \\
5f.6d &   3H*        &   5 &     4489.64 &   62903.70  &  0.056 \\
5f.6d &   3G*        &   5 &    11276.80 &   56116.54  &  0.210 \\
5f.6d &   3H*        &   6 &     8436.82 &   58956.52  &  0.839 \\  
\end{tabular}			
\end{ruledtabular}
\end{table}

The electron bridge mechanism reduces lifetime of the nuclear transition by the factor  $1+R$. The half-life in Th IV was measured to be $T_{1/2} =
1400^{+600} _{-300}$ s \cite{48}. This half-life is practically not affected
by EB, $(1+R) \approx 1$  \cite{EB1}.   The electron bridge mechanism  in Th III for electrons initially in the ground state decreases the nuclear transition half-life  $1+R=2.6$ times, giving  $T_{1/2} =
540^{+230} _{-115}$. 
 For electrons in the $J=0$ metastable state we obtained  $1+R=6.85$ and  $T_{1/2} =
200^{+90} _{-45}$.

\section{Dependence of the nuclear transition frequency on the states of electrons}

{ The calculation of the shifts of the nuclear transition frequency due to interaction with atomic electrons is similar to that of the atomic field isotope shift, and the results can be expressed in terms of the field-shift constants $F$ of the electronic states involved. In Ref.~\cite{w-shift} we calculated the field shift constants $F$  for the ground and several excited states of Th and its ions. 
 Using values of $F$ for the ground states, we presented shifts of nuclear transition frequencies for systems with different numbers of electrons.  
 
 Here, we consider a different situation in which the total number of electrons is fixed, but their electronic states are different.  We calculated $F$ for the  $J=0$  metastable state of Th~III, considered in the present work,  and determined the difference between the nuclear transition frequencies in the present system and 
 Th~III in its electronic ground state. We obtained}

\[
F(^3{\rm P}_0)=-30~\mathrm{GHz/fm}^2.
\]
The corresponding nuclear transition frequency shift is given by
\begin{equation}\label{e:dw}
\Delta\omega_N =
\left[F(^3{\rm P}_0)-F({\rm GS})\right]
\delta\langle r^2\rangle,
\end{equation}
where $F({\rm GS})=-68~\mathrm{GHz/fm}^2$~\cite{w-shift} is the field-shift constant for the electronic ground state of Th~III, 
and $\delta\langle r^2\rangle=0.0112~\mathrm{fm}^2$
\cite{w-shift} 
is the difference between the mean-square nuclear charge radii of the
isomeric and ground states of $^{229}$Th. Substituting these values into Eq.~(\ref{e:dw}), we obtain $\Delta\omega_N=430$~MHz, 
corresponding to a relative shift of $\Delta\omega_N/\omega_N=2.1\times10^{-7}$.

It is also important to find nuclear frequency shift for the first excited electronic state of Th~III: $6d^2 \ ^3$F$_2$ at $E=63$~cm$^{-1}$.
This is exceptionally long-living state connected to the ground state only by very weak M2 and E3 transitions. 
It appears as final state in decay of some excited states.
For example, it is the final state in the dominant decay channel of S2 (see Sec.~\ref{s:EB}).

The field shift factor $F$ for this state  was calculated in Ref.~\cite{w-shift}, yielding  $F(^3\rm{F}_2)$ = -40~GHz/fm$^2$.
Replacing $F(^3{\rm P}_0$) in (\ref{e:dw}) by this value, we obtain  $\Delta\omega_N=314$~MHz,  and $\Delta\omega_N/\omega_N=1.6\times10^{-7}$.

Note that the direct  contribution of  the core electrons to $F$ cancels out  in the difference of the values  of $F$ for exited and ground states and was not presented here.  However,  the core polarization effect has been calculated and included.

\section{Conclusion}

We have investigated the electronic bridge (EB) process for excitation
of the $^{229}$Th nuclear isomer in Th III ions initially prepared
in the metastable electronic state with $J=0$, recently proposed
for a nuclear clock with suppressed electronic hyperfine interactions.
Using the CI+SD method combined with the random-phase approximation,
we have calculated the EB transition rates for several single- and
two-laser excitation schemes.

We find that the EB enhancement strongly depends on the choice of
the final electronic state. For schemes in which the ion returns
to its initial $J=0$ state, the enhancement factor is relatively
small, $\beta\sim 1$, because of the absence of resonant contributions.
In contrast, allowing the final electronic state to differ from
the initial state leads to strong resonant enhancement.
In particular, the small energy denominator
$\Delta=-72~\mathrm{cm}^{-1}$ produces enhancement factors
as large as $\beta\approx 5\times 10^3$.

We have also considered single-laser EB excitation schemes,
which have smaller enhancement factors but offer a simpler
excitation procedure. An important practical issue is returning
the ion to its initial metastable $J=0$ state after nuclear
excitation. We have identified a possible repumping mechanism
involving a laser-stimulated electronic transition.

Our results demonstrate that efficient resonant excitation of
the $^{229}$Th nucleus is possible starting from the metastable
$J=0$ state of Th III. They also reveal a trade-off between
maximizing the EB excitation rate and preserving the advantages
of the $J=0$ electronic state for nuclear-clock operation.
The proposed schemes may facilitate the development of
high-precision nuclear clocks and their applications to
tests of fundamental physics.

 In addition, we calculated the shift in the nuclear transition frequency resulting from the change in the electronic state from the ground state to the excited metastable states, dependence of the nuclear transition lifetime on the  state of electrons and probabilities of electron transitions excited by the nuclear decay.  
 
 The electron bridge mechanism  in Th III for electrons initially in the ground state decreases the nuclear transition half-life  $1+R=2.6$ times, giving  $T_{1/2} =
540^{+230} _{-115}$. 
 For electrons in the $J=0$ metastable state  $1+R=6.85$ and  $T_{1/2} =
200^{+90} _{-45}$.

\textit{Acknowledgments.} -- This work  was supported by the Australian Research Council Grant No.\ DP230101058.

\end{document}